\documentclass[pre,onecolumn,preprintnumbers,amsmath,amssymb,superscriptaddress]{revtex4-1}

\usepackage{amsmath}
\usepackage{hyperref}
\usepackage{graphicx}
\usepackage{amsfonts}
\usepackage{amsthm}
\usepackage{cases}
\usepackage{bm}

\usepackage{color}
\definecolor{Blue}{rgb}{0.00, 0.00, 1.00}
\definecolor{Red}{rgb}{1.00, 0.00, 0.00}
\definecolor{Green}{rgb}{0.00, 0.70, 0.00}

\hypersetup{
    colorlinks=true,       
    linkcolor=red,          
    citecolor=blue,        
    filecolor=magenta,      
    urlcolor=cyan           
}

\newcommand{\be}{\begin{equation}}
\newcommand{\ee}{\end{equation}}
\newcommand{\bea}{\begin{eqnarray}}
\newcommand{\eea}{\end{eqnarray}}

\newcommand{\beq}{\begin{equation}}
\newcommand{\eeq}{\end{equation}}
\newcommand{\beqn}{\begin{eqnarray}}
\newcommand{\eeqn}{\end{eqnarray}}

\begin{document}

\title{Effusion of stochastic processes on a line}

\author{David S. \surname{Dean}}
\affiliation{Univ. Bordeaux, CNRS, LOMA, UMR 5798, F-33400, Talence, France}
\affiliation{Team MONC, INRIA Bordeaux Sud Ouest, CNRS UMR 5251, Bordeaux INP, Univ. Bordeaux, F-33400, Talence, France}
\author{Satya N. \surname{Majumdar}}
\affiliation{LPTMS, CNRS, Univ. Paris-Sud, Universit\'e Paris-Saclay, 91405 Orsay, France}
\author{Gr\'egory \surname{Schehr}}
\affiliation{Sorbonne Universit\'e, Laboratoire de Physique Th\'eorique et Hautes Energies, CNRS UMR 7589, 4 Place Jussieu, 75252 Paris Cedex 05, France}
\date{\today}
\begin{abstract} 

We consider the problem of leakage or effusion of an ensemble of independent 
stochastic processes from a region where they are initially randomly distributed. 
The case of Brownian motion, initially confined to the left half line with
uniform density and leaking 
into the positive half line is an example which has been extensively studied in the 
literature. Here we derive new results for the average number and variance of the 
number of leaked particles for arbitrary Gaussian processes initially
confined to the negative half line and also derive its  
joint two-time probability distribution, both for the
annealed and the quenched initial conditions. For the annealed case,
we show that the two-time joint distribution
is a bivariate Poisson 
distribution. We also discuss the role of correlations in
the initial particle positions on the 
statistics of the number of particles on the positive half line. We show that the strong 
memory effects in the variance of the particle number on the 
positive real axis for Brownian particles, seen in recent studies, persist 
for arbitrary Gaussian processes and also at the level of two-time correlation functions.

\end{abstract}

\maketitle
\section{Introduction}

A classical exercise in statistical mechanics courses is the study of effusion, the 
process by which a gas leaks from a container through a hole which is small with 
respect to the mean free path of the gas molecules. The one dimensional version of 
this problem is clearly relevant to the problem of dispersion in pore-like channels 
which appear in porous media such as zeolites \cite{KR92}. The dynamics of 
dispersion in one dimensional channels, especially in out of equilibrium situations 
when there are nonzero currents, has been intensively studied not just due
to their applications in one dimensional geometry, but also as a test bed
to develop general formalisms for non-equilibrium statistical mechanics~\cite{BD04,BSGJL05,BSGJ06,PM08,ADLV08,DG09,PS02,KM12}. 

Among this class of one dimensional problems, a particularly appealing one concerns 
the problem of effusion or contamination. First let us consider, for simplicity, 
the one-sided effusion problem where a set of stochastically evolving 
particles (interacting or noninteracting) {are initially confined to the 
negative half line with uniform density $\rho$. Then at $t=0$ this constraint is 
released}, allowing the particles to move on the full line. A natural quantity of 
interest is the net integrated current $Q(t)$ up to time $t$ through the origin $O$.
Clearly, for this one-sided initial condition, $Q(t)$ is just the number
of particles $N_+(t)$ located to the right of the origin at time $t$ and
originated from the left of the origin at time $t=0$.
Evidently $Q(t)$ is a random variable that depends
on the stochastic trajectories of the particles as well as
their initial conditions: what can we say about the probability 
distribution of $Q(t)$? How does this distribution depend on the 
initial correlations present in the initial condition?  
For noninteracting Brownian particles and also for particles undergoing symmetric
simple exclusion process, the distribution of $Q(t)$ was first computed 
in Ref.~\cite{DG09} using a variety of analytical tools. 
For noninteracting particles, a direct probabilistic approach was
used recently in Ref.~\cite{BMRS20} that allowed to compute the
current distribution for more general single particle dynamics, going beyond the
simple diffusion. This includes, for instance, 
run-and-tumble particles (RTP)~\cite{BMRS20} and more recently, stochastically
resetting particles~\cite{DHMMRS23}.

The one-sided effusion problem can be easily 
generalised to a two-sided case where both sides of the origin are initially occupied.
In the two sided case,
let us denote by $N_{\pm}(t)$ the number of particles that are to
the right (left) of the origin at time $t$, originating from the left (right) of the
origin at $t=0$. In this case, the net integrated current through the
origin up to time $t$ is simply the difference, $Q(t)=N_+(t)-N_{-}(t)$.
For noninteracting particles, the two random variables $N_+(t)$ and $N_-(t)$ are 
independent and thus the total
current $Q(t)$ can be easily determined from the independent studies
of the right (or left) effusion problems, where the particles
are initially located to the left (or right) of the origin.
The two sided problem has recently 
been revisited in the context of number fluctuations in ion channels and membrane 
pores~\cite{MAR21}. The statistics corresponding to two infinite half lines is 
relevant to the early time behavior for finite ion channels or pores, before finite 
size effects set in.

The results for $N_{\pm}(t)$ in the {\it noninteracting} problem can also be used
to study a certain number of observables in an {\it 
interacting} problem, namely the problem of elastically colliding stochastic 
processes. The simplest example corresponds to the case where point Brownian particles 
reflect off each other after each collision on a line, the so called
single file diffusion. The consequence of this reflection 
is that the order of the particles is maintained, i.e., 
if we label the particles say from $-M$ to $M$ on a finite 
segment of the real line, the label $m$ of a particle remains invariant
under the dynamics. Now consider a noninteracting problem
of Brownian particles where two particles simply cross each other upon
encounter. We can again label the particles in the noninteracting system and 
implement an effective reflection by interchanging the labels 
of two noninteracting particles whenever they cross each other. For instance, 
if particle with label $0$ is crossed from the right by the particle labelled $+1$, then the 
particle with label $+1$ is relabelled $0$ and the particle with
label $0$ is relabelled $+1$. This 
relabelling after each crossing leads to an interacting system of elastically 
colliding particles, and obviously can be generalised to arbitrary 
continuous-time stochastic processes, going beyond the simple Brownian motion.

For example, a much studied observable in the context of single file diffusion, 
is the position $Y(t)$ at time $t$ of a tagged or tracer particle, say with label $m=0$,
measured with respect to some fixed point in space. We can set the initial 
location $Y(0)=0$ without
any loss of generality. The tagged particle is
caged between its two neighbours and is well known to exhibit sub-diffusion at late times,
$Y(t)\sim t^{1/4}$~\cite{Harris65,Spitzer70,Richards77,Arratia83,Liggett83}.
The link with the two sided 
noninteracting effusion problem can be established~\cite{KMS14},
formalising the original discussion in~\cite{DGL85}, 
by considering the coarse-grained density field that plays the central role
in macroscopic fluctuation theory~\cite{BDG01}. 
Let $\rho(x,t)$ represent the particle density field which is identical
in both the interacting and noninteracting problem via the relabelling scheme. 
Then the number of particles
to the left of the $0$-th tagged particle in the interacting
problem must remain the same at all times (since the labelling is invariant in time) 
and hence
\begin{equation}
\int_{-\infty}^{Y(t)} \rho(x,t)\, dx = \int_{-\infty}^{0} \rho(x,0)\, dx \, .
\label{left}
\end{equation}
One can then express Eq. (\ref{left}) as {(see e.g. \cite{KMS14,RS23})}
\begin{equation}
\int_{0}^{Y(t) }\rho(x,t)\, dx= \int_{-\infty}^{0} \rho(x,0)\, dx-
\int_{-\infty}^{0} \rho(x,t)\, dx= N_L(0)-N_L(t) \, ,
\label{sflink}
\end{equation}
where $N_L(t)$ denotes the number of particles to the left of the origin at time 
$t$. Clearly, using the definitions introduced above, we have $N_L(t)=N_L(0) - 
N_+(t) + N_-(t)$ and so $N_L(0)-N_L(t)= N_+(t) - N_-(t)=Q(t)$. Now, as we expect $Y(t)$ 
to become large at late times, we can apply the law of large numbers to the density 
distribution \cite{DGL85} in the integral in Eq. (\ref{sflink}) to find
\begin{equation}
Y(t)\, \overline\rho \simeq N_+(t)-N_-(t)\, ,
\end{equation}
where $\overline\rho$ is the mean particle density which we assume is the same 
throughout the system. We thus find that at the level of {\rm typical} fluctuations
\begin{equation}
Y(t) \simeq \frac{N_+(t)-N_-(t)}{ \overline\rho}\, .
\label{link}
\end{equation}
Hence, by studying the statistics of $N_{\pm }(t)$ in the noninteracting
problem, one can infer the typical statistics of a tracer particle in the interacting
system (where particles reflect off each other upon collision). Note that
the discussion above is very general, and does not assume any specific type of dynamics.
Hence the result in Eq. (\ref{link}) holds for any elastically colliding stochastic
processes, including as special cases the single file diffusion and the
symmetric random average process on a line that also preserves the ordering of
particles~\cite{RM01}.
Furthermore, the statistics of the tracer displacement $Y(t)$ in the single file
diffusion is known to be equivalent to the statistics of height fluctuations
in $(1+1)$-dimensional Edwards-Wilkinson interfaces~\cite{BKK83,MB91,GMGB07} or
the position of a tagged monomer in a rouse polymer chain~\cite{GRT13}. Thus,
the statistics of $N_{\pm}(t)$ in the noninteracting effusion problem can
be directly related to these problems as well.

In addition to the stochastic evolution in time, the initial condition
also plays a crucial role on the statistics of the central objects $N_{\pm}(t)$
in the noninteracting effusion problem. 
Consider for simplicity, the
right effusion problem where all the particles are initially located to the
left of the origin.
There are two sources of randomness in $N_{+}(t)$, one coming from the noisy stochastic
evolution of the trajectories and another from the randomness of the initial conditions.
In analogy with disordered systems, the ``annealed"
case corresponds to averaging the distribution of $N_{\pm }(t)$ {\it simultaneously}
over the noise history and the initial condition. In contrast, in the
``quenched" case, one averages over the noise history for a fixed {\it typical}
initial configuration, e.g., when the particles are initially placed in 
a regular, equispaced configuration with 
spacing $1/\rho$ to the left of the origin. 
It has been shown that for uncorrelated Poissonian initial condition,
the annealed distribution of $N_+(t)$ is purely Poissonian at all $t$ for
any dynamics~\cite{BMRS20}. However, 
the distribution of $N_+(t)$ for the quenched case is highly 
nontrivial~\cite{DG09,BMRS20}.
For example, the large deviation function characterizing the tail of the
quenched distribution of $N_+(t)$ undergoes a third order phase transition
for stochastically resetting dynamics (both Brownian and RTP)~\cite{DHMMRS23}.
No signature of this phase transition is found in the annealed large deviation
of $N_+(t)$~\cite{DHMMRS23}. 
For a certain class of non-Poissonian initial conditions with a uniform density, 
the effect of initial conditions persists in the variance of the current even at
late times~\cite{LB13,CK17,BJC22}.

Most studies on the effusion problem, in noninteracting systems discussed above,
were focused on the one-time distribution of $N_+(t)$. A natural generalization 
is to ask: what can one say about the joint statistics of $N_+(t_1)$ and $N_+(t_2)$
at two different times $t_1$ and $t_2$? The natural temporal correlation of
the underlying dynamics of a single particle, coupled with the crossing statistics of 
the origin by multiple particles,
makes this two-time joint distribution of $N_+(t_1)$ and $N_+(t_2)$ nontrivial,
even for noninteracting particles. The goal of this paper is to
study and present exact results for this two-time distribution of $N_+(t_1)$ and $N_+(t_2)$ for 
noninteracting particles, but with very general dynamics and general initial 
conditions, and both for the annealed and the quenches cases. In particular,
the single particle dynamics in our noninteracting system is modeled
by a generic zero mean Gaussian process, which includes as special cases,
Brownian motion, fractional Brownian motion, 
thermalized underdamped Brownian motion etc. A certain number of 
our results can be directly transcribed to provide results for 
the tagged particle dynamics in the interacting problem of reflecting Gaussian processes  
via the link given in Eq. (\ref{link}) above. A brief summary of our results are
given below.

\vskip 0.5 truecm

{\noindent \bf Full two-time distribution of particles in a region $A$.} 
We first consider the annealed right effusion problem with the particles 
independently distributed initially on the left of the origin with 
uniform density $\rho$. Let $N_{A}(t_1)$ and $N_{A}(t_2)$ denote respectively 
the number of particles which are in a subset $A$ of the
real axis at times $t_1$ and~$t_2$.  
We compute exactly the joint probability distribution of $N_{A}(t_1)$ and $N_{A}(t_2)$
(rather its generating function) by averaging over the initial positions.
The statistics for $N_A(t_1)$ at a single time $t_1$ is 
Poissonian, and this result is well known for the case $A=\mathbb R^+$. Here we 
show that the two-time statistics are given by the bivariate Poisson distribution 
\cite{CI80}. The bivariate Poisson distribution has received little attention in the 
physics literature compared to the Poisson distribution whose occurrences are 
numerous. In the statistics literature, the bivariate Poisson distribution has 
notably been used to model the number of goals in football matches \cite{GKMS18}. The results we derive are general and the parameters of the bivariate Poisson 
distribution depend only on the average values $\mu(t_1)=\overline{\langle 
N_A(t_1)\rangle}$, $\mu(t_2)= \overline{\langle N_A(t_2)\rangle}$ and the annealed 
two-point function
\begin{equation}
C_{\rm an}(t_1,t_2)= \overline{\langle N_A(t_2)N_A(t_1)\rangle}- 
\overline{\langle N_A(t_2)\rangle}\ \ \overline{\langle N_A(t_1)\rangle},
\label{eqcan}
\end{equation}
where $\langle \cdots\rangle$ denotes an average over the stochastic trajectories for 
$t>0$ and $\overline{ \cdots}$ denotes the average over the distribution of the 
initial conditions of the particles.

We also consider the quenched problem where 
one averages the logarithm of the generating function
over the initial condition and then re-exponentiate the result~\cite{DG09,BMRS20}.
This quenched generating 
function allows us to compute the quenched two-time correlation function
\begin{equation}
C_{\rm qu}(t_1,t_2)=\overline{\langle N_A(t_2)N_A(t_1)\rangle}- 
\overline{\langle N_A(t_2)\rangle\langle N_A(t_1)\rangle}\, . 
\label{eqcq}
\end{equation}
However, the full quenched probability distribution function of $N_A(t)$ at two times 
does not seem to have a 
simple form. In the second term in Eq. (\ref{eqcq}),
one considers the thermal averages $\langle N_A(t_1)\rangle$
and $\langle N_A(t_2)\rangle$ both for the same initial condition,
multiply them and then average the product over initial conditions.
This averaging over the initial condition thus bears a similarity to the
averaging over the disorder in spin glasses via using the 
Edwards-Anderson order parameter~\cite{EA75}.

\vskip 0.5 truecm

{\noindent \bf The special case with $A={\mathbb R}^+$:}
We next focus on the most well studied example of $A={\mathbb R}^+$ 
and compute explicitly the one and two-time functions mentioned above for $N_{{\mathbb 
R}^+}= N_+(t)$. This computation is fully explicit for the full two-time distribution 
for the annealed case, and for moments up to order $2$ for the quenched case. We 
also give explicit results for different types of single
particle processes such as for Brownian motion, fractional Brownian motion, thermalized 
underdamped Brownian motion (which includes ballistic particles with a Maxwell 
Boltzmann distribution velocity as a certain limit, the Jepsen gas problem 
\cite{DJ65}) etc.

\vskip 0.5 truecm

{\noindent \bf The effect of correlated initial disorder on two-time correlation 
functions:} Beyond the difference between quenched and annealed averages various 
authors have discussed the role of initial conditions on the variance of $N_+(t)$.
This has been discussed 
mainly in the context of single file Brownian motion 
\cite{LB13,SD15,KMS14,BJC22} using macroscopic fluctuation theory valid
for general diffusive systems. 
Similar strong dependence on initial conditions has also been remarked much earlier
in one dimensional fluctuating interface models~\cite{KKMCBS97}
and also in the context of random average process on a line~\cite{RM01}. 
More recently, in Ref.~\cite{BJC22}, it was shown
that for a class of correlated initial conditions in the
effusion problem of diffusive systems, the variance of
$N_+(t)$ depends on the initial condition through the 
Fano-factor or the 
compressibility of the 
initial configuration. Here we generalize this result in two ways: it 
is extended to arbitrary Gaussian processes and also to the two-time connected 
correlation function. The strong dependence of the annealed two-time connected 
correlation function on this initial Fano-factor is seen, while the quenched 
two-time connected correlation 
function is, as expected on physical grounds, independent of the Fano-factor.

\vskip 0.5 truecm

The rest of the paper is organized as follows. In Section \ref{GF}, we provide
a general formalism to compute the generating function of the two-time distribution
of $N_A(t)$ in the right effusion problem in the noninteracting system, 
valid for arbitrary subset $A$ of the real line and for arbitrary single 
particle dynamics. The annealed and the quenched cases are discussed respectively
in subsections \ref{Annealed} and \ref{Quenched}. In Section \ref{Gaussian_processes} 
we obtain explicit results by applying the general formalism
to the example where the subset $A$ corresponds to the
positive half line and the single particle dynamics is governed
by a generic Gaussian process. Several specific examples are then
discussed in subsection \ref{Examples}. Section \ref{Initial_Condition}
discusses how the correlations present in the initial conditions
affect the two-time correlations of $N_A(t)$ in both the annealed
and the quenched cases. 
Finally, we conclude with a
summary and a discussion in Section \ref{Summary}.

\section{General formalism}
\label{GF}

In this section, we first develop a general formalism to compute the two-time
correlation function of the number of particles in a given subset in one dimension,
for a system of noninteracting particles but with arbitrary single-particle
dynamics. We start with a single particle whose position at time $t$ is denoted
by $X_t$ and it starts from the initial position $X_0=x_0$.
Let $A$ be a general 
subset of $\mathbb R $. For instance, $A$ can be a single interval or a
collection of disjoint intervals on the real line.
Let $I_A(x)$ be the indicator function of this subset, i.e.,
$I_A(x)=1$ if $x\in A$ and $I_A(x)=0$ if $x\in A^c$ where $A^c$ is the complement
of $A$.

We want to compute the joint probability distribution function (JPDF) of 
$I_A(X_{t_1})$  and  $I_A(X_{t_2})$ at two times $t_1<t_2$ and, 
initially, for a fixed value of $X_0$. For this, it is convenient to
consider the generating function of the JPDF 
\begin{equation}
G_2(z_1,z_2,x_0,t_1,t_2) =\Big\langle z_1^{I_A(X_{t_1})} z_2^{I_A(X_{t_2})}\Big\rangle \, ,
\label{genf_0}
\end{equation}
where the subscript $'2'$ in $G_2$ indicates the two-time function. Since the
indicator function takes values $0$ or $1$,
it is straightforward to see that 
\begin{equation}
G_2(z_1,z_2,x_0,t_1,t_2) = \Big\langle \left[1 +
(z_1-1)\,I_A(X_{t_1})\,I_{A^c}(X_{t_2}) +(z_2-1)\,I_A(X_{t_2})\,I_{A^c}(X_{t_1}) +
(z_1\,z_2-1)\, I_A(X_{t_1})\,I_A(X_{t_2})\right] \Big\rangle \, ,
\label{two-time_genf.1}
\end{equation}
where $I_{A^c}(x)=1-I_A(x)$ denotes the indicator function of the complement $A^c$. 
There are many ways of representing the above expression 
but we note that the combination of indicator function products used in 
Eq. (\ref{two-time_genf.1}) is such that at most one of
the last three terms can be non-zero. 
This decomposition will make it straightforward to interpret our results later for 
an ensemble of independent particles in terms of the bivariate Poisson distribution. 

If the JPDF of $X_{t_1}$ and $X_{t_2}$, conditioned on $X_0=x_0$, is denoted by
$p(x_2,t_2;\,x_1,t_1|x_0)$ then we can write
\begin{eqnarray}
\langle I_A(X_{t_2})I_{A^c}(X_{t_1})\rangle &=& \int_A dx_2 \int_{A^c} dx_1\, 
p(x_2,t_2;x_1,t_1|x_0) \nonumber \\
\langle I_{A^c}(X_{t_2})I_{A}(X_{t_1})\rangle &=& \int_{A^c} dx_2 
\int_{A} dx_1\, p(x_2,t_2;x_1,t_1|x_0)\nonumber \\
\langle I_A(X_{t_2})I_{A}(X_{t_1})\rangle &=& \int_A dx_2 \int_{A} dx_1\, 
p(x_2,t_2;x_1,t_1|x_0) \;.
\label{p_cond.1}
\end{eqnarray}

Now consider an ensemble of $N$ independent particles with positions $X_{it}$
at time $t$, each with initial condition $X_{i0}=x_{i0}$.  
The number of particles in the interval $A$ at time $t$ is simply given by
\begin{equation}
N_A(t) = \sum_{i=1}^N I_A(X_{it})\, .
\end{equation}
Clearly $N_A(t)$ is a random variable with two sources of randomness:
the first is due to the stochasticity or the noise of the evolving trajectories  
and the second is the randomness coming from the initial positions of the particles, i.e.,
the disorder associated with the initial conditions. 
We now discuss the two types of initial disorder averaging 
commonly treated in the literature, namely the annealed and the quenched cases.

\begin{figure}[t!]
\begin{center}
  \includegraphics[scale=0.2]{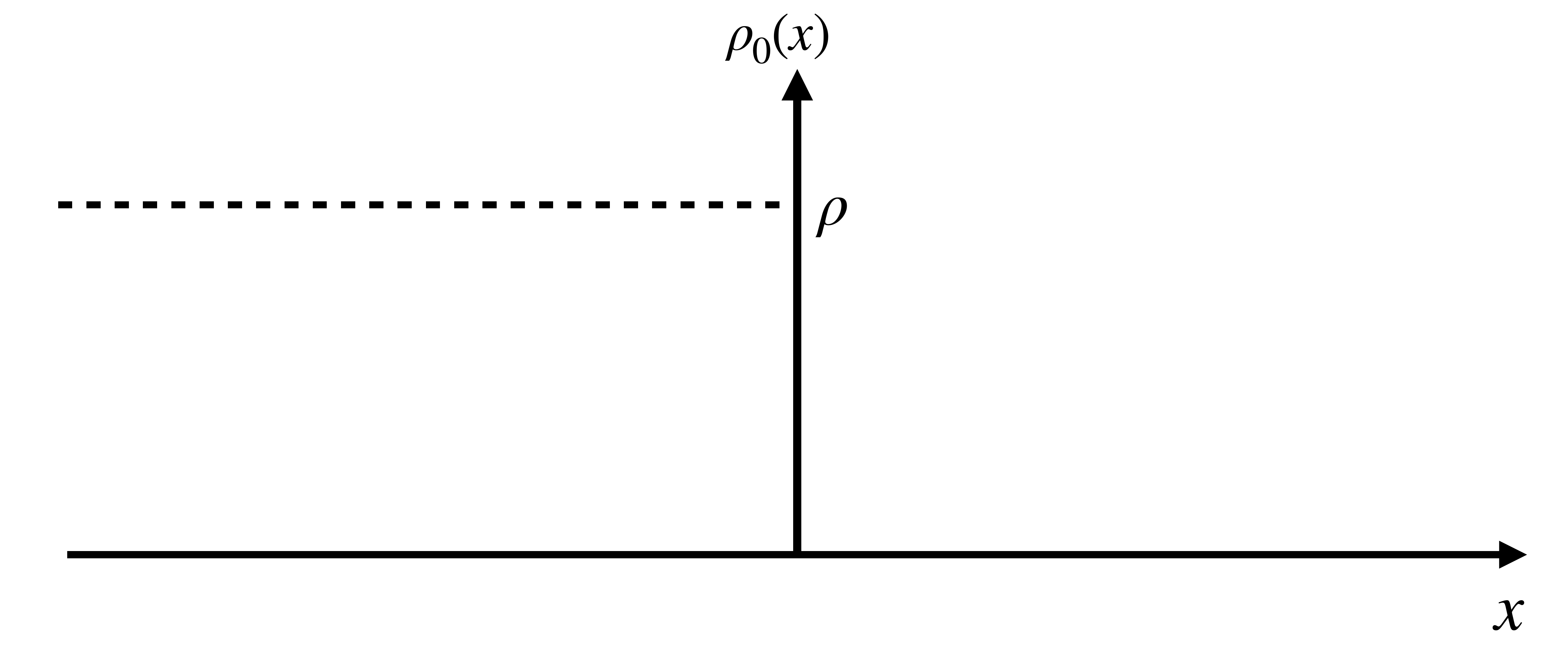} 
 \caption{Step like configuration of initial average particle density.}
\label{fig.init}
\end{center}
\end{figure}

\subsection{Annealed initial condition}
\label{Annealed}

The annealed joint generating function averaged over the initial conditions is
\begin{equation}
g_{2,\rm an}(z_1,z_2,t_1,t_2) =\overline{\Big\langle z_1^{N_A(t_1)}\,
 z_2^{N_A(t_2)}\Big\rangle} = \overline{ \prod_{i=1}^N G_2(z_1,z_2,x_{i0},t_1,t_2)]} \;,
 \end{equation}
where $\overline{\cdots}$ indicates averaging over the initial condition. Taking 
derivatives with respect to $z_1$ and $z_2$ produces joint moments of the type 
$\overline{\langle N^k_A(t_2)N^{k'}_A(t_1)\rangle}$ where both averages over 
stochastic trajectories and the initial conditions are carried
out simultaneously.

For simplicity, we will consider the right effusion problem, i.e., we assume that 
initially at $t=0$ all the particles are confined in $[-L,0]$ to the left of the 
origin. In the computation of 
annealed averages, we assume further that each 
particle at $t=0$ is distributed independently and uniformly over $[-L,0]$.
Eventually, we will take the limit $L\to \infty$, $N\to \infty$ with the density 
$\rho=N/L$ kept fixed on the negative half line (see Fig.~\ref{fig.init}).
In this limit, the initial particle 
number locally has a Poisson distribution with mean $\rho$. The initial condition 
thus contains fluctuations which will clearly affect the statistics of how the 
particles leak out of the negative half line as time progresses.
Carrying out the average over the initial condition then gives
\begin{equation}
g_{2,\rm an}(z_1,z_2,t_1,t_2) =\left[ \frac{1}{L}\int_{-L}^0 dx_0\, 
G_2(z_1,z_2,x_{0},t_1,t_2)\right]^N \, .
\label{gan.1}
\end{equation}
We now substitute $G_2$ from Eq. (\ref{two-time_genf.1}) into Eq. (\ref{gan.1}),
take the limit $N\to \infty$, $L\to \infty$ with the ratio $\rho=N/L$ fixed and
use Eq. (\ref{p_cond.1}) to obtain
\begin{equation}
g_{2, \rm an}(z_1,z_2,t_1,t_2)
= \exp\left[\lambda_{10}(t_1,t_2)\,(z_1-1) +\lambda_{01}(t_1,t_2)\,(z_2-1) +
\lambda_{11}(t_2,t_1)\,(z_2\,z_1-1)\right]\, ,
\label{gan.2}
\end{equation}
with
\begin{eqnarray}
\lambda_{01}(t_1,t_2)&=&\rho\int_{-\infty}^0 dx_0 \int_{A} dx_2 \int_{A^c} 
dx_1\, p(x_2,t_2;x_1,t_1|x_0) \;,\nonumber \\
 \lambda_{10}(t_1,t_2)&=& \rho\int_{-\infty}^0 dx_0 \int_{A^c} dx_2\ \int_{A} dx_2\, 
p(x_2,t_2;x_1,t_1|x_0) \;, \nonumber \\
\lambda_{11}(t_1,t_2) &=&\rho \int_{-\infty}^0 dx_0 \int_A dx_2 \int_{A} dx_1\, 
p(x_2,t_2;x_1,t_1|x_0) \, .
\label{lambda_def}
\end{eqnarray}

We now recall the definition of the bivariate Poisson distribution~\cite{CI80}.
First consider a single Poisson random variable $N_1$ with parameter $\lambda_1$.
The generating function of its distribution has the simple form
\begin{equation}
g_1(z)= \langle z^{N_1}\rangle =e^{-\lambda_1}\, \sum_{N_1=0}^{\infty} 
\frac{z^{\lambda_1\,N_1}}{N!}= e^{\lambda_1\, (z_1-1)}\, .
\label{genf_Poisson.1}
\end{equation}
Two random variables $N_1$ and $N_2$ have a bivariate Poisson distribution if they can 
be written as $N_1 = M_1+ M_{12}$ and $N_2 = M_2 + M_{12}$ where $M_1$ and $M_2$ are 
independent Poisson random variables with parameters $\lambda_1$ and $\lambda_2$, 
while the common term $M_{12}$ is another independent Poisson random variable with 
parameter $\lambda_{12}$. From this definition we see that the generating function 
for the joint distribution of $N_1$ and $N_2$ is given by 
\begin{equation}
g_2(z_1,z_2) =\langle z_1^{N_1}z_2^{N_2}\rangle = \langle z_1^{M_1}\rangle\, 
\langle z_2^{M_2}\rangle\, \langle (z_1\, z_2)^{M_{12}}\rangle=
 \exp\left[\lambda_1\,(z_1-1) + \lambda_2\,(z_2-1) +\lambda_{12}\, (z_2\,z_1-1)\right]\, ,
\label{genf_Poisson.2}
\end{equation}
where we used the independence of $M_1$, $M_2$ and $M_{12}$.
By formally inverting the generating function, one can represent
the joint probability distribution $P(N_1,\,N_2)$ as \cite{CI80,GKMS18}
\begin{equation}
P(N_1,\,N_2)=e^{-\lambda_1-\lambda_2 -\lambda_{12}}\, \frac{\lambda_1^{N_1}
\lambda_2^{N_2}}{N_1! N_2!}\, \sum_{n=0}^{{\rm min}(N_1,N_2)} n!\, 
\begin{pmatrix}N_2 & \\ n\end{pmatrix}\begin{pmatrix}N_1 & 
\\ n\end{pmatrix}\, \left(\frac{\lambda_{12}}{\lambda_1\lambda_2}\right)^n\, .
\end{equation}
However for most practical purposes the joint generating function $g_2(z_1,z_2)$
in Eq. (\ref{genf_Poisson.2}) is simpler and more useful. 

Comparing Eqs. (\ref{gan.2}) and (\ref{genf_Poisson.2}) we see that
the generating function of the annealed two-time JPDF of $N_A(t_1)$ and $N_A(t_2)$
has the bivariate Poisson form with parameters $\lambda_1\equiv \lambda_{10}(t_1,t_2)$,
$\lambda_2\equiv \lambda_{01}(t_1,t_2)$ and $\lambda_{12}\equiv \lambda_{11}(t_1,t_2)$
given in Eq. (\ref{lambda_def}). Consequently, we can represent the random variables
$N_A(t_1)$ and $N_A(t_2)$ as
\begin{equation}
N_A(t_1) \equiv M_1 + M_{12}, \quad\, N_A(t_2) \equiv M_2 + M_{12}
\end{equation}
where $M_{1}$, $M_2$ and $M_{12}$ are independent Poisson
variables with parameters $\lambda_{10}(t_1,t_2)$,  
$\lambda_{01}(t_1,t_2)$, and
$\lambda_{11}(t_2,t_1)$ respectively. Note that this result is valid for
arbitrary interval $A$ on the line. Physically, $M_{12}$
represents the number of particles that are present in $A$ at both
time $t_1$ and $t_2$. The quantity $M_1$ represents the number of particles
in $A$ and $t_1$ that are not present in $A$ at time $t_2$. 
Similarly, $M_2$ represents the number of particles in $A$ at time $t_2$
that were not present in $A$ at time $t_1$.

The generating function for the distribution of $N_A(t_1)$ at a single time 
$t_1$ can be simply obtained by marginalizing 
\begin{equation}
g_{1,\rm an}(z_1,t_1)=g_{2,\rm an}(z_1,z_2=1,t_1,t_2) =  
\exp\left(\mu(t_1)\,(z_1-1)\right) \, ,
\label{marginal.1}
\end{equation}
with
\begin{equation}
\mu(t_1)=\lambda_{10}(t_1,t_2) + \lambda_{11}(t_1,t_2)= 
\rho\, \int_{-\infty}^0 dx_0 \int_A dx_1\, p(x_1,t_1 |x_0)\, ,
\label{marginal.2}
\end{equation}
where $p(x_1,t_1|x_0)$ is simply the probability density to reach $x_1$ at time $t_1$,
starting at $x_0$ at $t=0$. In deriving the last equality in Eq.~(\ref{marginal.2}),
we used the definitions of $\lambda_{10}(t_1,t_2)$ and $\lambda_{11}(t_1,t_2)$
in Eq. (\ref{lambda_def}) and the fact that $\int_{-\infty}^{\infty} dx_2\, 
p(x_2,t_2; x_1,t_1|x_0)= p(x_1,t_1|x_0)$. As expected, 
$\mu(t_1)$ in Eq. (\ref{marginal.2}) depends only on $t_1$. 
The annealed distribution of $N_A(t_1)$ at a single time $t_1$ is thus obviously 
Poisson~\cite{BMRS20}
\begin{equation}
{\rm Prob.}[N_A(t_1)=n]= p(n,t_1)=e^{-\mu(t_1)}\, \frac{\mu(t_1)^n}{n!}\, ,
\end{equation}
with the mean and the variance given by
\begin{equation}
\overline{\langle N_{A}(t_1)\rangle}\Big|_{\rm an} = \mu(t_1); \quad\quad 
{\rm Var}[N_A(t_1)]\Big|_{\rm an}= \overline{\langle N^2_{A}(t_1)\rangle}- 
\left[\overline{\langle N_{A}(t_1)\rangle}\right]^2 = \mu(t_1) \, .
\label{mean_var.an}
\end{equation}
Finally, it is easy to see that  
\begin{eqnarray}
\lambda_{10}(t_1,t_2)&=& \mu(t_1)- \lambda_{11}(t_1,t_2) \;, \\
\lambda_{01}(t_1,t_2)&=& \mu(t_2)- \lambda_{11}(t_1,t_2) \;,
\end{eqnarray}
using which we can obtain an alternative representation of the two-time generating function
\begin{equation}
g_{2,\rm an}(z_1,z_2,t_1,t_2) = \exp\left[\mu(t_1)\,(z_1-1)+ \mu(t_2)\,(z_2-1) +
\lambda_{11}(t_1,t_2)\,(z_1-1)(z_2-1)\right]\, .
\label{alt_g2}
\end{equation}
Note that this result can also be derived directly starting from Eq. (\ref{genf_0}).

The annealed two-time correlation function  
can be easily computed from the representation
$N_A(t_1)= M_1 +M_{12}$ and $N_A(t_2)= M_2+ M_{12}$ and
using the independence of the Poisson variables $M_1$, $M_2$ and $M_{12}$
(alternatively by taking derivatives of the generating function).
One gets for the connected correlation function
\begin{eqnarray}
C_{\rm an}(t_1,t_2)&=& \overline{\langle N_A(t_1)N_A(t_2)\rangle}- \overline{\langle N_A(t_1)\rangle}
\ \overline{\langle N_A(t_2)\rangle}\nonumber \\
&=&\langle (M_1 + M_{12})((M_2 + M_{12})\rangle -\langle (M_1 + M_{12})\rangle\langle 
(M_2 + M_{12})\rangle \nonumber \\
&=& \langle M_{12}^2\rangle - \langle M_{12}\rangle^2 = \lambda_{11}(t_1,t_2)\, ,
\label{an_two-time.1}
\end{eqnarray}
where $\lambda_{11}(t_1,t_2)$ is given in Eq. (\ref{lambda_def}) in terms
of the two-time propagator of a single particle $p(x_2,t_2;x_1,t_1|x_0)$.
Note that the result in Eq. (\ref{an_two-time.1})
holds for any single-particle dynamics and any interval $A$ on the line.
This is one of the main general results of this paper.

A number of other results are also immediate. For instance, 
the probability that no particles are in $A$ at time $t$ is given by
\begin{equation}
P(N_A(t)=0) = \exp\left[-\mu(t)\right] \;.
\end{equation}
The probability distribution of $N_A(t_2)$ conditioned on $N_A(t_1)=0$  is Poisson with
\begin{equation}
P(N_A(t_2)=n| N_A(t_1)=0) = \exp(-\mu(t_2)+ \lambda_{11}(t_1,t_2))\frac{(\mu(t_2)-\lambda_{11}(t_2,t_2))^n}{n!}\;,
\end{equation}
while  the probability distribution of $N_A(t_1)$ conditioned on $N_A(t_2)=0$  is Poisson with
\begin{equation}
P(N_A(t_1)=n| N_A(t_2)=0) = \exp(-\mu(t_1)+ \lambda_{11}(t_1,t_2))\frac{(\mu(t_1)-\lambda_{11}(t_2,t_1))^n}{n!}\;.
\end{equation}

\subsection{Quenched initial condition}
\label{Quenched}

The quenched generating function is defined as
\begin{equation}
g_{2,\rm qu}(z_1,z_2,t_1,t_2) = \exp\left(\,\, 
\overline{\ln\left[\langle z_1^{N_A(t_1)}\,z_2^{N_A(t_2)}\rangle\right]}\,\, \right)\, .
\label{gfq.1}
\end{equation}
Using the definition in Eq. (\ref{genf_0}), it follows that
\begin{equation}
\overline{\ln[\langle z_1^{N_A(t_1)}z_2^{N_A(t_2)}\rangle]}=\overline{ \sum_{i=1}^N \ln G_2(z_1,z_2,x_{i0},t_1,t_2)]}
\label{qgenf.1}
\end{equation}
where $G_2$ is given in Eq. (\ref{two-time_genf.1}). Performing the average over disorder (with
initial positions chosen independently and uniformly over $[-L,0]$) and using 
Eq. (\ref{p_cond.1}) we get
\begin{eqnarray}
&&\overline{\ln[\langle z_1^{N_A(t_1)}z_2^{N_A(t_2)}\rangle]}= 
\frac{N}{L}\int_{-L}^0 dx_0\ln\left[1+ (z_1-1)\int_{A^c} dx_2 \int_{A} dx_1\, p(x_2,t_2;x_1,t_1|x_0)\right. \nonumber \\
&+&\left.
(z_2-1)\int_{A} dx_2 \int_{A^c} dx_1\, p(x_2,t_2;x_1,t_1|x_0) +(z_2z_1-1)\int_{A} dx_2 \int_{A} dx_1\, 
p(x_2,t_2;x_1,t_1|x_0)\right] \, .
\label{qgenf.2}
\end{eqnarray}
Finally, substituting this result on the right hand side (rhs)
of Eq. (\ref{gfq.1}) and taking the limit $N\to \infty$, $L\to \infty$ with $\rho=N/L$ fixed, we get
\begin{eqnarray}
&&g_{2,\rm qu}(z_1,z_2,t_1,t_2) = 
\exp\left(\rho\, \int_{-\infty}^0 dx_0\ln\left[ 1+ (z_1-1)\int_{A^c} dx_2 \int_{A} dx_1\, 
p(x_2,t_2;x_1,t_1|x_0)\right.\right.\nonumber \\
&+&\left.\left.
(z_2-1)\int_{A} dx_2 \int_{A^c} dx_1\, p(x_2,t_2;x_1,t_1|x_0) 
+(z_2z_1-1)\int_{A} dx_2 \int_{A} dx_1\, p(x_2,t_2;x_1,t_1|x_0)\right]\right)\, .
\label{qgenf.3}
\end{eqnarray}

From this exact quenched generating function $g_{2,\rm qu}$, 
we can derive explicitly several quantities
of interest. For example, the generating function for the one-time distribution
can be obtained by setting $z_2=1$ in Eq. (\ref{qgenf.3}). One gets
\begin{equation}
g_{1,\rm qu}(z_1,t_1) = g_{2,\rm qu}(z_1,z_2=1,t_1,t_2)= 
\exp\left(\rho\, \int_{-\infty}^0 dx_0\, \ln\left[1+ (z_1-1) \int_{A} dx_1\, p(x_1,t_1|x_0)\right]\right) \, .
\label{qu_one_point.1}
\end{equation}
Consequently, the quenched average value of $N_A(t_1)$ at time $t_1$ can be computed as
\begin{equation}
\overline{\langle N_{A}(t_1)\rangle}\Big|_{qu}=\frac{\partial \ln[g_{1,\rm qu}(z_1,t_1)]}{\partial z_1}\Big|_{z_1=1}= \mu(t_1)=
\rho\, \int_{-\infty}^0 dx_0 \int_A dx_1 \, p(x_1,t_1 |x_0)\,,
\label{qmean.1}
\end{equation}
which coincides with the annealed average in Eq. (\ref{mean_var.an}). 

Next we calculate the two-time quenched correlation function defined in Eq. (\ref{eqcq}). Starting from
the expression of $g_{2,\rm qu}$ in Eq. (\ref{qgenf.3}) and taking derivatives with respect to both $z_1$ and $z_2$,
one gets
\begin{equation}
C_{\rm qu}(t_1,t_2)= \overline{\langle N_A(t_2)N_A(t_1)\rangle}-
\overline{\langle N_A(t_2)\rangle\langle N_A(t_1)\rangle}
=\frac{\partial^2 \ln[g_{2\, \rm qu}(z_1,z_2,t_1,t_2)]}{\partial 
z_1\partial z_2}\Big|_{z_1=z_2=1}=\lambda_{11}(t_1,t_2) -\lambda_{11}^*(t_1,t_2) \, ,
\label{q_two-time.1}
\end{equation}
where $\lambda_{11}(t_1,t_2)$ is defined in Eq. (\ref{lambda_def}) and
\begin{equation}
\lambda_{11}^*(t_1,t_2)=\rho\, \int_{-\infty}^0 dx_0  \int_{A} dx_1 p(x_1,t_1|x_0)\, \int_{A} dx_2\, p(x_2,t_2|x_0)
= \rho\, \int_{-\infty}^0 dx_0\, \langle I_{A}(X_{t_2})\rangle\, \langle I_A(X_{t_1})\rangle \, .
\label{qlambda11_def}
\end{equation}
Noting from Eq. (\ref{an_two-time.1}) that $C_{\rm an}(t_1,t_2)=\lambda_{11}(t_1,t_2)$, we find the interesting relation between the quenched and the annealed two-time correlations
\begin{equation}
C_{\rm an}(t_1,t_2)- C_{\rm qu}(t_1,t_2)= \lambda_{11}^*(t_1,t_2) \ge 0\, .
\label{anqu.1}
\end{equation}
Thus the annealed two-time correlator is larger than its quenched counterpart, signifying a larger 
fluctuation in the former case. 

To summarize this section, we have presented a general formalism
for the noninteracting right effusion problem where the particles are uniformly distributed
on the left half of the origin at $t=0$ with density $\rho$.
This formalism allows us to compute the
joint distribution of the number of particles $N_A(t_1)$ and $N_A(t_2)$ 
in an arbitrary interval $A$, at two different times $t_1$ and $t_2$. 
The generating function for the joint distribution of $N_A(t_1)$ and $N_A(t_2)$, 
both for the annealed and the quenched cases, are given respectively 
in Eqs. (\ref{alt_g2}) and (\ref{qgenf.3}).
In the annealed case, the generating function in Eq. (\ref{alt_g2}) has a bivariate Poisson form 
fully characterized by just two quantities $\mu(t)$ and $\lambda_{11}(t_1,t_2)$
defined respectively in Eqs. (\ref{marginal.2}) and (\ref{lambda_def}).
The annealed two-time correlation function, from Eq. (\ref{an_two-time.1}), is simply
$C_{\rm an}(t_1,t_2)= \lambda_{11}(t_1,t_2)$.
In the quenched case, the generating function
in Eq. (\ref{qgenf.3}) has a less explicit form. However, the expression for the two-time
quenched correlation function in Eq. (\ref{q_two-time.1}) is explicit:
$C_{\rm qu}(t_1,t_2)= \lambda_{11}(t_1,t_2)- \lambda_{11}^*(t_1,t_2)$ with
$\lambda_{11}^*(t_1,t_2)$ given in Eq. (\ref{qlambda11_def}). 
These results for the noninteracting
right effusion problem are very general
and hold for any interval $A$ and any single-particle dynamics. 
To determine the full joint distribution at two times in the
annealed case and the two-time correlator in the quenched case, 
we need to just evaluate three quantities given the interval $A$, namely 
\begin{eqnarray}
\mu(t) & = & \rho\, \int_{-\infty}^0 dx_0 \int_A dx\, p(x,t|x_0)\, , \label{mu.1} \\
\lambda_{11}(t_1,t_2) &=& \rho\, \int_{-\infty}^0 dx_0 \int_A dx_2 \int_{A} dx_1\,
p(x_2,t_2;x_1,t_1|x_0)\, , \label{l11.1} \\
\lambda_{11}^*(t_1,t_2) &=& \rho\, \int_{-\infty}^0 dx_0  
\int_{A} dx_1\, p(x_1,t_1|x_0)\, \int_{A} dx_2\, p(x_2,t_2|x_0)\, . \label{l11*.1}
\end{eqnarray}
All three quantities depend essentially only on 
the two-time propagator $p(x_2,t_2;x_1,t_1|x_0)$ of
the underlying single-particle dynamics, which is the key quantity.
For any interval $A$ and any single-particle dynamics for which
the three integrals above can be computed
explicitly, one would have the explicit two-time correlators and the 
full two-time joint distribution in the annealed case.
In the next section, we demonstrate how this can be carried
out for Gaussian processes and provide a few interesting examples.

\begin{figure}[t!]
\begin{center}
  \includegraphics[width=0.4\linewidth]{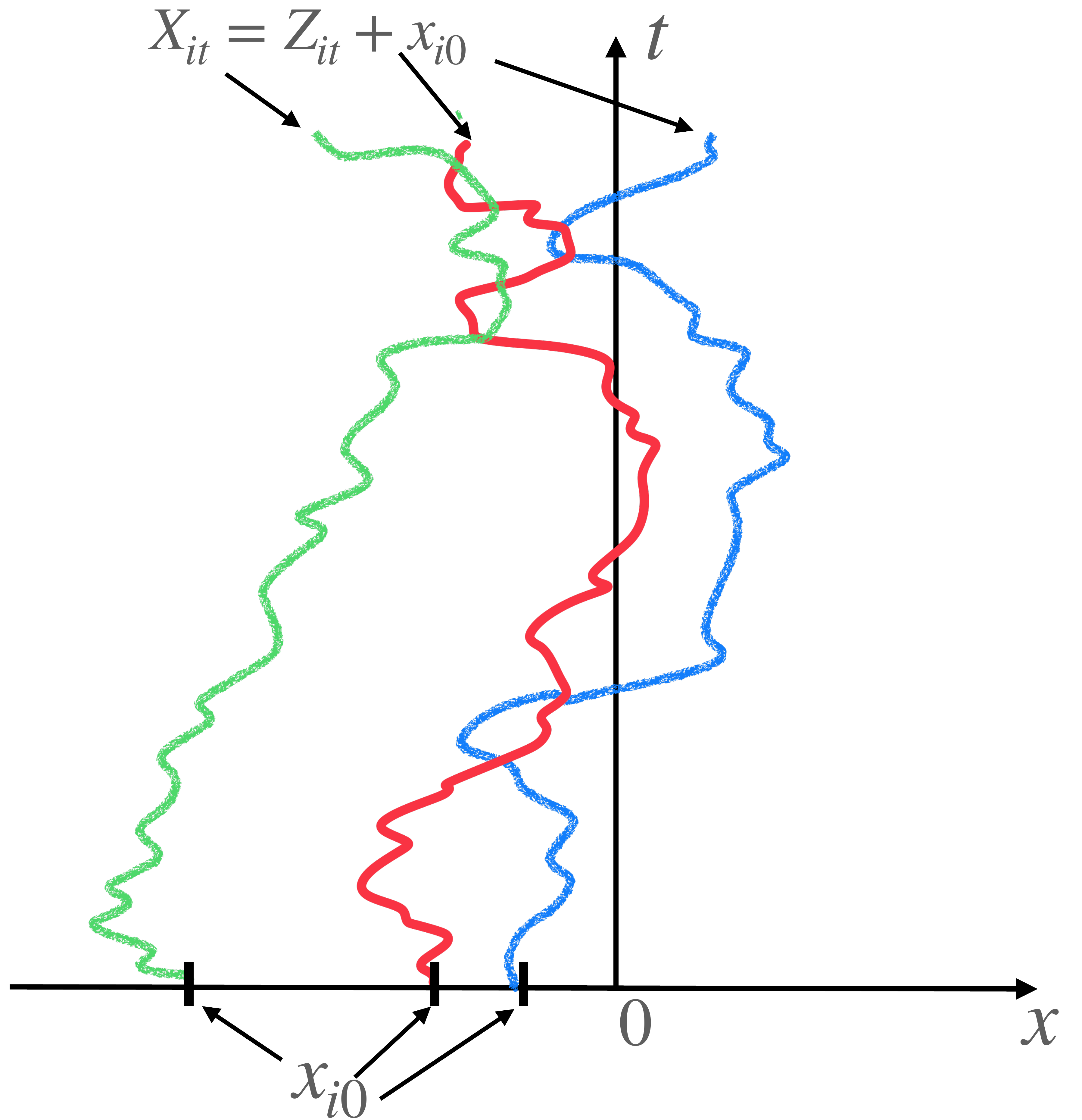} 
 \caption{{Schematic representation of the stochastic trajectories} $X_{it}=Z_{it}+x_{i0}$ of particles
labelled by $i$, started from 
{$X_{i0}=x_{0}$} or equivalently $Z_{i0}=0$ at time $t=0$. 
When a particle crosses the origin from the left it contributes to the number of 
particles $N_+(t)$, but no longer contributes once is crosses back to the left. }
\label{fig.schema}
\end{center}
\end{figure}

\section{Application of the general formalism: Gaussian processes}
\label{Gaussian_processes}

In this section, we focus on the right effusion problem
with the choice $A=\mathbb R^+$. In this case, $N_A(t)= N_+(t)$, i.e.,
the number of particles at time $t$ on the positive half axis (see Fig. 
\ref{fig.schema}). Furthermore,
we consider the single-particle dynamics to be governed by a generic Gaussian process, that 
includes, as special cases, the Brownian motion, fractional Brownian motion, 
noisy underdamped oscillator etc. For Gaussian processes, we show below
that all the three quantities $\mu(t)$, $\lambda_{11}(t_1,t_2)$ and
$\lambda_{11}^*(t_1,t_2)$ given respectively in Eqs. (\ref{mu.1}), (\ref{l11.1})
and (\ref{l11*.1}) can be computed explicitly.

We consider a single-particle dynamics where the position $X_t$ at time $t$,
starting at $x_0$ initially, can be written as $X_t=Z_t+x_0$, where $Z_t$
is a generic, zero mean, Gaussian process which is completely characterized by
its two-time correlation function
\begin{equation}
\langle Z_t\, Z_{t'} \rangle = c(t,t')\ .
\label{corr.1}
\end{equation}
Since the interval $A$ is just $[0,\infty)$, the three key quantities
in Eqs. (\ref{mu.1}), (\ref{l11.1}) and (\ref{l11*.1}) can be expressed as
\begin{eqnarray}
\mu(t) & = & \rho\, \int_{-\infty}^0 dx_0\, \langle \theta(Z_t+x_0)\rangle\, , 
\label{mu.2} \\
\lambda_{11}^*(t_1,t_2) &=& 
\rho\, \int_{-\infty}^0 dx_0\,
\langle \theta(Z_{t_2}+x_0)\rangle\,
\langle \theta(Z_{t_1}+x_0)\rangle  \, , \label{l11*.2} \\
\lambda_{11}(t_1,t_2)& =& 
\rho\, \int_{-\infty}^0 dx_0\, \langle \theta(Z_{t_2}+x_0)
\, \theta(Z_{t_1}+x_0)\rangle\, ,
\label{l11.2}
\end{eqnarray}
where $\theta(x)$ is the Heaviside step function: $\theta(x)=1$ if $x>0$
and $\theta(x)=0$ if $x<0$.

Let us start by computing $\mu(t)$ in Eq. (\ref{mu.2}). 
To evaluate $\langle \theta(Z_t+x_0)\rangle$, we just need
the single-time propagator of the zero mean Gaussian process $Z_t$ which is simply
\begin{equation}
P(z,t)= \frac{1}{\sqrt{2\, \pi\, c(t,t)}}\, \exp\left(- \frac{z^2}{2\, c(t,t)} \right)\, .
\label{1time_prop.1}
\end{equation}
Consequently,
\begin{equation}
\langle \theta(Z_t+x_0)\rangle= \int_{-x_0}^{\infty} P(z,t)\, dz  
=\frac{1}{2}\, {\rm erfc}\left(-\frac{x_0}{\sqrt{2\, c(t,t)}}\right)\, ,
\label{theta_int.1}
\end{equation}
where ${\rm erfc}(z)= (2/\sqrt{\pi})\, \int_z^{\infty} e^{-u^2}\, du$ is the
complementary error function. Substituting Eq. (\ref{theta_int.1}) on the rhs
of Eq. (\ref{mu.2}) and performing the integral over $x_0$ gives
\begin{equation}
\mu(t) = \frac{\rho}{2}\, \int_0^\infty dx_0\ 
{\rm erfc}\left(\frac{x_0}{\sqrt{2 c(t,t)}}\right)=
\rho\, \sqrt{\frac{c(t,t)}{2\pi}}\, .
\label{eint}
\end{equation}

Similarly, $\lambda_{11}^*(t_1,t_2)$ in Eq. (\ref{l11*.2}) also requires
only the integral in Eq. (\ref{theta_int.1}) and
can be evaluated explicitly to give
\begin{eqnarray}
\lambda_{11}^*(t_1,t_2)& = & \frac{ \rho}{4}\, \int_{-\infty}^0 dx_0\, 
{\rm erfc}\left(\frac{-x_0}{\sqrt{2\, c(t_2,t_2)}}\right)\, 
{\rm erfc}\left(\frac{-x_0}{\sqrt{2 c(t_1,t_1)}}\right) \nonumber \\
&=& \frac{\rho}{2\sqrt{2\pi}}\left(\sqrt{c(t_1,t_1)}+ \sqrt{c(t_2,t_2)}-
\sqrt{c(t_1,t_1) + c(t_2,t_2)}\right) \;.
\label{l*}
\end{eqnarray}

It then rests to compute $\lambda_{11}(t_1,t_2)$ in Eq. (\ref{l11.2}) which
is more nontrivial since it
requires computing the two-time function 
$\langle \theta(Z_{t_2}+x_0)
\, \theta(Z_{t_1}+x_0)\rangle$ for a zero mean Gaussian process $Z_t$.
The computation of this two-time correlator for a generic
zero mean Gaussian process $Z_t$ has been widely discussed in the literature 
and in particular for the case $x_0=0$, where Slepian \cite{SLE62} derived 
an explicit formula for it, and also for the three-time correlator
$\langle \theta(Z_{t_3})
\, \theta(Z_{t_2})\, \theta(Z_{t_1})\rangle$.
The extension of these explicit results to
higher order correlation functions of Heaviside theta functions for a Gaussian
process is not known and
remains an outstanding problem in the study of persistence \cite{BMS13}.

In order to compute $\lambda_{11}(t_1,t_2)$ in Eq. (\ref{l11.2}),
we then need an extension of Slepian's formula for $x_0=0$ to a nonzero $x_0$ for
the two-time correlator $\langle \theta(Z_{t_2}+x_0)
\, \theta(Z_{t_1}+x_0)\rangle$. This can be done as follows. First, to compute
this two-time correlator, we need the two-time joint 
distribution $P(z_2,t_2; z_1,t_1)$
of the zero mean Gaussian process $Z_t$, since 
\begin{equation}
\langle \theta(Z_{t_2}+x_0)
\, \theta(Z_{t_1}+x_0)\rangle= \int_0^{\infty}dz_2\, \int_0^{\infty} dz_1\, 
P(z_2,t_2;z_1,t_1) \, .
\label{Z2time.1}
\end{equation}
The two-time JPDF of $Z_t$ can
be most conveniently expressed in the Fourier space
\begin{eqnarray}
\int_{-\infty}^{\infty} dk_2 \int_{-\infty}^{\infty} dk_1\, e^{i\, k_2\, z_2}\,
e^{i\, k_1\, z_1}\, P(z_2,t_2;z_1,t_1) &= &\Big\langle e^{i\, k_2\, Z_{t_2} + 
i\, k_1\, Z_{t_1}}\Big\rangle \nonumber \\
&= & \exp\left(- \frac{1}{2}\left[ k_2^2\, c(t_2,t_2) + 2\, k_2\, k_1\, c(t_1,t_2)
+ k_1^2\, c(t_1,t_1)\right]\right)\, ,
\label{Z2t.1}
\end{eqnarray}
with $c(t_1,t_2)=c(t_2,t_1)$.  

To compute the double integral in Eq. (\ref{Z2time.1}), it is also convenient to work
in the Fourier space. We start with the following integral representation
of delta function
\begin{equation}
\delta(x) = \frac{1}{2\pi}\int_{-\infty}^\infty dk\ \exp(ikx),
\label{delta_int.1}
\end{equation}
along with that of the Heaviside theta function
\begin{equation}
\theta(x) = \int_0^\infty dx'\, \delta(x-x') = 
\int_0^\infty dx' \frac{1}{2\pi}\, \int_{-\infty}^\infty dk\, \exp(ik(x-x'))= 
-i\int_{-\infty}^\infty\frac{dk}{2\pi k}\exp(ik x),
\end{equation}
where the integration contour for $k$ is taken just below the real axis to ensure convergence. This thus yields the identity
\begin{equation}
\theta(Z_t + x_0) =-i\int_{-\infty}^\infty \frac{dk}{2\pi k}\exp(i\,k\, Z_t+i\,k\,x_0)\, .
\end{equation}
Using this identity twice and averaging, using Eq. (\ref{Z2t.1}), gives
\begin{eqnarray}
J(x_0,{t_1},{t_2})&= &\langle \theta(Z_{t_2} + x_0)\theta(Z_{t_1} + x_0)\rangle \nonumber \\  
&=& -\int_{-\infty}^\infty \int_{-\infty}^\infty\frac{dk_2\,dk_1}{4\pi^2 k_2\,k_1}
\exp\left(-\frac{1}{2}\,\left[k_2^2\, c(t_2,t_2) + 2\,k_2\, k_1\, 
c(t_1,t_2) + k_1^2\, c(t_1,t_1)\right]+i\,(k_2+k_1)\, x_0\right).
\label{def_Jx0}
\end{eqnarray}
Next we notice that 
\begin{equation}
\frac{\partial J(x_0,t_1,t_2)}{\partial c(t_1,t_2)}  
=\int_{-\infty}^\infty\int_{-\infty}^\infty \frac{dk_2\,dk_1}{4\pi^2}
\exp\left(-\frac{1}{2}\left[k_2^2\, c(t_2,t_2) + 2\,k_2\, k_1\, c(t_1,t_2) + 
k_1^2\, c(t_1,t_1)\right]+i(k_2+k_1)x_0\right) \, .
\end{equation}
This Gaussian integral can be carried out explicitly to give
\begin{equation}
\frac{\partial J(x_0,t_1,t_12)}{\partial c(t_1,t_2)}  =
\frac{1}{2\,\pi\, \sqrt{c(t_2,t_2)\,c(t_1,t_1)-c^2(t_1,t_2)}}\,
\exp\left(-\frac{[c(t_2,t_2)+ c(t_1,t_1)-2\,c(t_1,t_2)]}{2\,
[c(t_2,t_2)\,c(t_1,t_1)-c^2(t_1,t_2)]}\, x_0^2
\right) \, .
\label{j1}
\end{equation}
When $x_0=0$, it is easy to integrate this equation with respect to $c(t_1,t_2)$
and use the fact that when $c(t_1,t_2)=0$, i.e., for the uncorrelated case,
one simply has $J(0,t_1,t_2)=1/4$ (this is just the probability
that two independent zero mean Gaussian variables are both positive).
One then recovers the classical result by Slepian~\cite{SLE62}, 
\begin{equation}
\langle \theta(Z_{t_2})\theta(Z_{t_1})\rangle =J(0,t_1,t_2)= 
\frac{1}{2\pi}\,\sin^{-1}\left(\frac{c(t_1,t_2)}{\sqrt{c(t_2,t_2)c(t_1,t_2)}}\right)+
\frac{1}{4}\, .
\end{equation}

However, for $x_0$ nonzero, it is not easy to integrate Eq. (\ref{j1}) with respect to
$c(t_1,t_2)$. Fortunately, to calculate $\lambda_{11}(t_1,t_2)$ in Eq. (\ref{l11.2}),
we do not need to evaluate $J(x_0,t_1,t_2)$ for a fixed $x_0$, but rather we need 
the integral of
$J(x_0,t_1,t_2)$ over $x_0\in [-\infty,0]$. It turns out that integrating
first Eq. (\ref{j1}) over $x_0$ and then integrating over $c(t_1,t_2)$ simplifies
the computation. 
Performing the integral of Eq. (\ref{j1}) over $x_0$ first gives 
\begin{equation}
\int_{-\infty}^0 dx_0\frac{\partial J(x_0,t_1,t_2)}{\partial c(t_1,t_2)} = 
\frac{1}{2\sqrt{2\pi}\sqrt{c(t_2,t_2)+c(t_1,t_1) - 2\, c(t_1,t_2)}}\, .
\end{equation}
Now we can easily integrate with respect to $c(t_1,t_2)$ to get
\begin{equation}
\lambda_{11}(t_1,t_2)= \int_{-\infty}^0 dx_0\, J(x_0, t_2,t_1)=
\frac{\rho}{2\sqrt{2\pi}}\left[\sqrt{c(t_2,t_2)}+\sqrt{c(t_1,t_1)}- 
\sqrt{c(t_2,t_2)+c(t_1,t_1) - 2\, c(t_1,t_2)}\right] + A_0\, ,
\label{l11.3}
\end{equation}
where $A_0$ is an integration constant. To fix $A_0$, we again examine
the uncorrelated case when $c(t_1,t_2)=0$. In this case, we have
\begin{eqnarray}
\lambda_{11}(t_1,t_2)\Big|_{c(t_1,t_2)=0}
&= &\rho\, \int_{-\infty}^0 dx_0\, \langle \theta(Z_{t_2}+x_0)\theta(Z_{t_1}+x_0)
\rangle\Big|_{c(t_1,t_2)=0}\, .
\label{uncorr.1}
\end{eqnarray}
However, in the absence of correlation, the product $\langle \theta(Z_{t_2}+x_0)\theta(Z_{t_1}+x_0)
\rangle$ decouples and comparing to Eq. (\ref{l*}), we immediately see that
\begin{equation}
\lambda_{11}(t_1,t_2)\Big|_{c(t_1,t_2)=0}= \lambda_{11}^*(t_1,t_2)=\frac{\rho}{2\sqrt{2\pi}}\left(\sqrt{c(t_1,t_1)}+ \sqrt{c(t_2,t_2)}-
\sqrt{c(t_2,t_2) + c(t_1,t_1)}\right)\, .
\label{uncorr.2}
\end{equation}
Setting $c(t_1,t_2)=0$ in Eq. (\ref{l11.3}) and using Eq. (\ref{uncorr.2}) immediately fixes the
integration constant $A_0=0$. Noting that the annealed two-time correlator
is simply $\lambda_{11}(t_1,t_2)$ [see Eq. (\ref{an_two-time.1})] and the quenched 
two-time correlator is $\lambda_{11}(t_1,t_2)-\lambda_{11}^*(t_1,t_2)$ from 
Eq. (\ref{q_two-time.1}), we then get for generic Gaussian processes 
\begin{eqnarray}
C_{\rm an}(t_1,t_2)&= &\lambda_{11}(t_1,t_2)= \frac{\rho}{2\sqrt{2\pi}}\,
\left[\sqrt{c(t_2,t_2)}+\sqrt{c(t_1,t_1)}-
\sqrt{c(t_2,t_2)+c(t_1,t_1) - 2\, c(t_1,t_2)}\right]\, , \label{la11} \\ 
C_{\rm qu}(t_1,t_2)&=& \lambda_{11}(t_1,t_2)- \lambda_{11}^*(t_1,t_2)
=\frac{\rho}{2\sqrt{2\pi}}\left[\sqrt{c(t_1,t_1) + c(t_2,t_2)}- 
\sqrt{c(t_2,t_2)+c(t_1,t_1) - 2\, c(t_1,t_2)}\right] \, . \label{lq11}
\end{eqnarray}  

Another interesting observable is the instantaneous current $j(0,t)$ at time $t$ through
the origin in the right effusion problem. When the single-particle dynamics is
pure diffusion, one can read off this current from the diffusion equation as
$j(0,t)=-D\, \partial_x \rho(x,t)|_{x=0}$ where $\rho(x,t)$ is the density field. 
However, for general Gaussian processes (including non-Markovian ones for which
one can not write a simple Fokker-Planck equation), it is not immediately clear
how to compute the statistics of the instantaneous current $j(0,t)$. Here,
the approach used above can be used to compute the mean instantaneous current,
as well as the two-time current correlation functions, both in the annealed
and the quenched cases. To proceed, we note the simple fact that
\begin{equation}
j(0,t) =\frac{dN_+(t)}{dt} \, .
\label{j0t_def}
\end{equation}
This immediately gives, both in the annealed and the quenched cases,
\begin{equation}
\overline{\langle j(0,t)\rangle}= \frac{d\mu(t)}{dt}\, ,
\label{mean_j0t.1}
\end{equation}
and hence from Eq. (\ref{eint})
\begin{equation}
\overline{\langle j(0,t)\rangle}=\frac{\rho}{\sqrt{2\pi}}\frac{d}{dt}\sqrt{ c(t,t)}\, .
\label{mean_j0t.2}
\end{equation}
The annealed two-time current-current correlation function can be expressed
in terms of the annealed correlator $C_{\rm an}(t_1,t_2)$ in Eq. (\ref{la11}) as
\begin{eqnarray}
G_{\rm an}(t_1,t_2)&= &\overline{\langle j(0,t_1)j(0,t_2)\rangle}- 
\overline{\langle j(0,t_1)\rangle}\ \overline{\langle j(0,t_2)\rangle} \nonumber \\
&=&\frac{\partial^2}{\partial t_2\partial t_1}C_{\rm an}(t_1,t_2)
= \frac{\rho}{2\sqrt{2\pi}}\frac{\partial^2}{\partial t_2\partial t_1}
\left[\sqrt{c(t_2,t_2)}+\sqrt{c(t_1,t_1)}- 
\sqrt{c(t_2,t_2)+c(t_1,t_1) - 2\, c(t_1,t_2)}\right]\, .
\label{cu_corr_an.1}
\end{eqnarray}
Similarly the quenched two-time current correlator 
can be expressed in terms of $C_{\rm qu}(t_1,t_2)$ in Eq. (\ref{lq11}) as
\begin{eqnarray}
G_{\rm qu}(t_1,t_2)&= & \overline{\langle j(0,t_1)\rangle}\overline{\langle 
j(0,t_2)\rangle}- \overline{\langle j(0,t_1)\rangle\langle j(0,t_2)\rangle} \nonumber \\
&=&\frac{\partial^2}{\partial t_2\partial t_1}C_{\rm qu}(t_1,t_2)
= \frac{\rho}{2\sqrt{2\pi}}\frac{\partial^2}{\partial t_2\partial t_1}
\left[\sqrt{c(t_1,t_1) + c(t_2,t_2)}- \sqrt{c(t_2,t_2)+c(t_1,t_1) - 
2\, c(t_1,t_2)}\right]\, .
\label{cu_corr_qu.1}
\end{eqnarray}

\subsection{Specific Examples}
\label{Examples}

In this subsection we consider some specific examples of Gaussian processes
that appear quite commonly and provide explicit results for the annealed
and the quenched cases.

\vskip 0.5 truecm

{\noindent \bf Brownian and Fractional Brownian motion.} The fractional Brownian
motion (fBM) is a commonly studied Gaussian process with zero mean and correlator
\begin{equation}
c(t_1,t_2) = D_H\, \left(t_1^{2H} + t_2^{2H}- |t_2-t_1|^{2H} \right)\, ,
\label{fbm_corr.1}
\end{equation}
where $0<H<1$ is the Hurst exponent and $D_H$ the fractional diffusion constant.
The case $H=1/2$ corresponds to the ordinary Brownian motion, where Eq. (\ref{fbm_corr.1})
reduces to the standard Brownian correlator
\begin{equation}
c(t_1,t_2)= 2\, D\, {\rm min}(t_1,t_2)\, ,
\label{brown_corr.1}
\end{equation}
with $D_{1/2}=D$ denoting the diffusion constant of a Brownian motion.
Substituting the correlator (\ref{fbm_corr.1}) in
Eq. (\ref{eint}), we get 
\begin{equation}
\mu(t)= \rho\,\sqrt{ \frac{D_H}{\pi}}\, t^{H} \, .
\label{mu_fbm.1}
\end{equation}
The annealed and the quenched correlators of $N_+(t)$, 
from Eqs. (\ref{la11}) and (\ref{lq11}),
are given by
\begin{eqnarray}
C_{\rm an}(t_1,t_2) &= & 
\frac{\rho\sqrt{D_H}}{2\sqrt{\pi}}\,\left(t_1^H+t_2{^H}- |t_2-t_1|^H\right)\, , \label{an_fbm.1}\\
C_{\rm qu}(t_1,t_2) &= & 
\frac{\rho\sqrt{D_H}}{2\sqrt{\pi}}\,\left(\sqrt{t_1^{2H}+ t_2^{2H}}- |t_2-t_1|^H\right)\, .
\label{qu_fbm.1}
\end{eqnarray}
For the Brownian case $H=1/2$ these results agree with those given in~\cite{LB13,SD15}.

We next consider the statistics of the instantaneous current $j(0,t)$ for the fBM.
Using Eq. (\ref{mean_j0t.2}) and $c(t_1,t_2)$ from Eq. (\ref{fbm_corr.1}), we get the mean instantaneous current
\begin{equation}
\overline{\langle j(0,t)\rangle}=\rho\, H\,\sqrt{\frac{D_H}{\pi}}\, t^{H-1}\, .
\label{mean_incurr_fbm.1}
\end{equation}
The annealed two-time current-current correlation function in Eq. (\ref{cu_corr_an.1})
gives (assuming $t_2>t_1$)
\begin{equation}
G_{\rm an}(t_1,t_2)=
\overline{\langle j(0,t_1)j(0,t_2)\rangle}- 
\overline{\langle j(0,t_1)\rangle}\ 
\overline{\langle j(0,t_2)\rangle}= 
-\frac{\rho\,\sqrt{D_H}\,H\,(1-H)}{2\,\sqrt{\pi}\, (t_2-t_1)^{2-H}} \, ,
\end{equation}
while in the quenched case we get
\begin{equation}
G_{\rm qu}(t_1,t_2)=\overline{\langle j(0,t_1) j(0,t_2)\rangle}- 
\overline{\langle j(0,t_1)\rangle\langle j(0,t_2)\rangle}= 
-\frac{\rho\,\sqrt{D_H}\,H\,(1-H)}{2\,\sqrt{\pi}\, (t_2-t_1)^{2-H}}-
\frac{\rho\,\sqrt{D_H}\,H^2}{2\,\sqrt{\pi} }\frac{t_1^{2H-1}\,t_2^{2H-1}}{(t_1^{2H} 
+t_2^{2H})^\frac{3}{2}}.
\end{equation}
We thus see that the current fluctuations in both the annealed and quenched cases 
are anti-correlated and furthermore this anti-correlation is more enhanced
in the quenched case compared to the annealed case. The 
negative correlation in the current fluctuations can be understood 
physically as follows.
If at a 
certain moment of time $j(0,t)$ has a spike, it signifies
the event that a particle has crossed the 
origin from left to right at that moment. However the particle must still be close to the 
origin for some time immediately after and typically recrosses back soon after 
crossing, leading to a dip in $j(0,t)$. This explains the anti-correlation. 
Interestingly, for the annealed current fluctuations, the amplitude
\begin{equation}
A(H) = \frac{\rho\, \sqrt{D_H}\,H\,(1-H)}{2\,\sqrt{\pi}} \;,
\end{equation}
as a function of the $H$ is maximal at $H=1/2$ corresponding to
the Brownian case and tends to zero as $H$ tends to $0$ or $1$. 
Both the annealed and quenched current correlators 
diverge at equal times due to the singular nature of the current. 
However, one should notice that the quantity
\begin{equation}
\overline{\langle j(0,t_1)\rangle\langle j(0,t_2)\rangle}- 
\overline{\langle j(0,t_1)\rangle}\, 
\overline{\langle j(0,t_2)\rangle} =
\frac{\rho\, \sqrt{D_H}\,H^2}{2\,\sqrt{\pi} }\frac{t_1^{2H-1}\,t_2^{2H-1}}{(t_1^{2H} +
t_2^{2H})^\frac{3}{2}},
\end{equation}
which measures the fluctuations of the trajectory averaged current 
(which is clearly smooth) with respect to the disorder in the initial conditions, 
exhibits positive correlations. Furthermore, due to the smoothing caused 
by the averaging over trajectories, this correlation function is 
finite at $t_2=t_1$ where it takes the value
\begin{equation}
\overline{\langle j(0,t_1)\rangle^2}- \left[\overline{\langle j(0,t_1)\rangle}\right]^2
=\frac{\rho\, \sqrt{D_H}\,H^2}{4\,\sqrt{2\pi}\, t_1^{2-H}} \, .
\end{equation}
This measure of the current fluctuations depends 
strongly on correlations in the initial conditions and 
we will revisit this question later in Section \ref{Initial_Condition}.

\vskip 0.5 truecm

\noindent {\bf Thermalized underdamped Brownian motion and the Jepsen gas.} Another 
example is where the process $Z_t=\int_0^t V_{t'}\, dt'$ is a 
thermalized underdamped Brownian motion (i.e.,
the physical Brownian motion) of 
mass $m$. Here the velocity $V_t=\frac{dZ_t}{dt}$ evolves via the Langevin equation
\begin{equation}
m\frac{dV_t}{dt}=-\gamma\, V_t + \sqrt{2\,\gamma\, k_B\,T}\, \eta(t),
\label{OU.1}
\end{equation}
where $\gamma$ is the friction coefficient, $T$ is the temperature and $k_B$
is the Boltzmann constant. The noise $\eta(t)$ is a zero mean Gaussian white
noise with correlator $\langle\eta(t)\eta(t')\rangle= \delta(t-t')$. 
We assume that the initial velocity $V_0$ is chosen
from the Maxwell-Boltzmann distribution: $P(V_0)\propto \exp[-k_B\, T\, m\, V_0^2/2]$.

Since Eq. (\ref{OU.1}) is linear and $V_0$ has a Gaussian distribution, $V_t$
is a zero mean Gaussian process in time and consequently $Z_t$ is also a zero mean
Gaussian process. 
By integrating Eq. (\ref{OU.1}) for $V_t$ and computing its correlator, one gets
$\langle V_{t_1}V_{t_2}\rangle= (T/m)\, e^{-\gamma\, |t_2-t_1)|/m}$,
where we used $\langle V_0^2\rangle= k_B T/m$. Consequently, the correlation function
for $Z_t= \int_0^t V_{t'}\, dt'$ is given by 
\begin{equation}
c(t_1,t_2) = \frac{k_B\,T }{m}\int_0^{t_2} ds_2\int_0^{t_1} ds_1\,
\exp\left(-\frac{|s_2-s_1|}{\tau} \right)\, ,
\label{OU_corr.1}
\end{equation}
where $\tau = m/\gamma$ is the relaxation time separating the 
ballistic short time regime from the diffusive long time regime. 
Explicitly at equal times, the correlation function is given by
\begin{equation}
c(t,t)= \frac{2\,k_B\,T\tau^2 }{m}\, 
\left(\frac{t}{\tau}-1 + e^{-\frac{t}{\tau}}\right)
\label{ctt_ou.1}
\end{equation}
and at different times (setting $t_2\ge t_1$) one has
\begin{equation}
c(t_1,t_2) = c(t_1,t_1) + \frac{k_B\, T\, \tau^2 }{m}\left(1+
e^{-\frac{t_2}{\tau}}-e^{-\frac{t_1}{\tau}}-e^{-\frac{t_2-t_1}{\tau}}\right)\, .
\label{ct1t2_ou.1}
\end{equation}
In the limit $t_1,\ t_2\ \gg \tau$ and $ t_2-t_1\ \gg \tau$,
one recovers the Brownian result (fBM with $H=1/2$ discussed above)
with diffusion constant $D=k_B T/\gamma$. However,
two other limits are quite interesting that we discuss below.

\vskip 0.3cm

\noindent {\it The limit $t_1,\ t_2\ \gg \tau$, but $t_2-t_1\ll \tau$}.
In this case, it follows from Eq. (\ref{ctt_ou.1}) that
$c(t,t)\approx 2\, D\, t$ with $D=k_B T/\gamma$ as expected.
Consequently, from Eq. (\ref{eint}), one gets
\begin{equation}
\mu(t)= \rho\, \sqrt{\frac{D\, t}{\pi}}\, .
\label{mu_lim.1}
\end{equation}
However, the correlator at different times in Eq. (\ref{ct1t2_ou.1})
turns out to be slightly more nontrivial 
\begin{equation}
c(t_1,t_2)\approx D\,(t_1+t_2) \, .
\label{corr_lim.1}
\end{equation}
Consequently, from Eqs. (\ref{la11}) and (\ref{lq11}) we get the
annealed and the quenched two-time correlators as
\begin{eqnarray}
C_{\rm an}(t_1,t_2) &\approx & \frac{\rho\, \sqrt{D}}{2\, \sqrt{\pi}}\, \left(\sqrt{t_1}+
\sqrt{t_2}\right)\, , \label{an_lim.1} \\
C_{\rm qu}(t_1,t_2) &\approx & \frac{\rho\, \sqrt{D}}{2\, \sqrt{\pi}}\, \sqrt{t_1+t_2}\, .
\label{qu_lim.1}
\end{eqnarray}

\vskip 0.3cm

\noindent {\it The Jepsen limit: $t_1,\ t_2\ \ll \tau$}. In this case, the
particles behave ballistically with initial velocity chosen from the
Maxwell-Boltzmann distribution. This is the so called Jepsen gas~\cite{DJ65}.
For this Jepsen gas, initially 
confined on the negative half line and 
leaking into the positive half line with increasing time, 
various single-time observables have been computed exactly, e.g.,
the position and the velocity distribution of the leader particle at 
time $t$~\cite{BM07}. Here we compute the two-time
correlators of the number of particles $N_+(t)$ on the right half for
this leaking Jepsen gas. 
In this limit, expanding the exponential on the rhs of Eq. (\ref{ctt_ou.1})
up to {$O(t^2)$} we get
\begin{equation}
c(t,t) \approx \frac{k_B\, T}{m}\, t^2 \, .
\label{ctt_jep.1}
\end{equation}
Consequently, from Eq. (\ref{eint}), we get that the mean increases linearly with time, i.e.,  
\begin{equation}
\mu(t) \approx \rho\, \sqrt{\frac{k_B\, T}{2\,\pi\, m}}\, t  \, .
\label{mu_jep.1}
\end{equation}
Similarly, the correlator at different times in Eq. (\ref{ct1t2_ou.1}) reduces simply to
\begin{equation}
c(t_1,t_2)\approx \frac{k_B\, T}{m}\, t_1\, t_2 \, .
\label{corr_jep.1}
\end{equation}
Consequently, the annealed and the quenched two-time correlators from Eqs. (\ref{la11})
and (\ref{lq11}) read (setting $t_2\ge t_1$)
\begin{eqnarray}
C_{\rm an}(t_1,t_2) &\approx & \rho\, \sqrt{\frac{k_B\, T}{2\,\pi\, m}}\, t_1
\label{an_jep.1} \\
C_{\rm qu}(t_1,t_2) &\approx & 
\frac{1}{2}\, \rho\, \sqrt{\frac{k_B\,T}{2\,\pi\, m}}\left[\sqrt{t_2^2+t_1^2}- 
t_2+ t_1\right] \, .
\label{qu_jep.1}
\end{eqnarray}

\section{Correlated initial conditions for Gaussian processes}
\label{Initial_Condition}

Here we examine the effect of initial conditions on $N_+(t)$ along the 
lines of Ref.~\cite{BJC22}. We will show how their analysis 
for diffusive processes at a single time can be extended to two-time 
quantities and also to arbitrary Gaussian processes 
going beyond the diffusion.
We recall the definition in Eq. (\ref{eqcan}) of the annealed two-time correlation function 
\begin{equation}
C_{\rm an}(t_1,t_2)=\overline{ \langle N_+(t_2)\,N_+(t_1)\rangle} - 
\langle \overline{N_+(t_2)\rangle}\ \ \overline{\langle N_+(t_1)\rangle}\, .
\label{ic_an.1}
\end{equation}
Following Ref.~\cite{BJC22} we decompose this as 
\begin{equation}
C_{\rm an}(t_1,t_2) =C_{\rm qu}(t_1,t_2)+ C_{\rm ic}(t_1,t_2)\, ,
\label{ic_decom.1}
\end{equation}
where 
\begin{equation}
C_{\rm qu}(t_1,t_2) =\overline{\langle N_+(t_2)\,N_+(t_1)\rangle} - 
\overline{\langle  N_+(t_2)\rangle \langle N_+(t_1)\rangle} \,
\label{ic_qu.1} 
\end{equation}
is the quenched correlation function defined in Eq. (\ref{eqcq}). The correlation function 
$C_{\rm ic}(t_1,t_2)$ is given by
\begin{equation}
C_{\rm ic}(t_1,t_2)= \overline{\langle  N_+(t_2)\rangle\, \langle N_+(t_1)\rangle}- 
\overline{\langle  N_+(t_2)\rangle }\ \ \overline{\langle N_+(t_1)\rangle}
\label{ic_mixed.1}
\end{equation}
and it measures correlations in the trajectory averaged values of 
$N_+(t)$ induced by the randomness in the initial conditions. 

We again assume that the single particle dynamics is governed by a 
zero mean Gaussian process,
i.e., the position of the $i$-th particle at time $t$ is simply $X_{it}=Z_{it}+x_{i0}$,
where $x_{i0}$ is the initial position of the $i$-th particle
and $Z_{it}$ is a zero mean Gaussian process. Then 
\begin{equation}
N_+(t)=\sum_{i=1}^N \theta(Z_{it}+x_{i0}) \,
\label{ic_nplus}
\end{equation}
where $Z_{it}$'s for different $i$ are uncorrelated, but the initial positions
$x_{i0}$'s for different $i$ are correlated.
Using this we find 
\begin{equation}
C_{\rm qu}(t_1,t_2)= \sum_{ij}\overline{\langle\theta(Z_{it_2}+x_{i0})\,
\theta(Z_{jt_1}+ x_{j0})\rangle-
\langle\theta(Z_{it_2}+x_{i0})\rangle\, \langle\theta(Z_{jt_1}+x_{j0})\rangle}\, .
\label{ic_qu.2}
\end{equation}
However, as the processes $Z_{it}$ are independent, averaging over the particle 
trajectories leaves only the diagonal terms nonzero in the above double sum, leading to
 \begin{equation}
C_{\rm qu}(t_1,t_2)= \sum_{i=1}^N\overline{\langle\theta(Z_{it_2}+x_{i0})\,
\theta(Z_{it_1}+x_{i0})\rangle-
\langle\theta(Z_{it_2}+x_{i0})\rangle\, \langle\theta(Z_{it_1}+x_{i0})\rangle} \, .
\label{ic_qu.3}
\end{equation}

We thus see that  $C_{\rm qu}(t_1,t_2)$ does not depend on correlations in the 
initial conditions. Assuming that the initial position $x_{i0}$ is distributed
uniformly over $[-L,0]$ we carry out the average $\overline{ \cdots}$
in Eq. (\ref{ic_qu.3}) and then take the limit $N\to \infty$, $L\to \infty$ with
$\rho=N/L$ fixed. We then find that $C_{\rm qu}(t_1,t_2)$ coincides with
the result in Eq. (\ref{lq11}) for independent initial conditions, namely
\begin{equation}
C_{\rm qu}(t_1,t_2)= \frac{\rho}{2\sqrt{2\pi}}\,
\left[\sqrt{c(t_1,t_1) + c(t_2,t_2)}- \sqrt{c(t_2,t_2)+c(t_1,t_1) - 
2\, c(t_1,t_2)}\right] \, .
\label{ic_qu.4}
\end{equation}
For the case of Brownian diffusion, where $c(t_1,t_2)= 2\, D\, {\rm min}(t_1,t_2)$,
this result in Eq. (\ref{ic_qu.4}) agrees
with Ref.~\cite{BJC22}.

In contrast to the quenched two-time correlator $C_{\rm qu}(t_1,t_2)$,
we now show that the correlator $C_{\rm ic}(t_1,t_2)$ in Eq. (\ref{ic_mixed.1})
does depend on the correlations in the initial conditions.
Taking average over trajectories in Eq. (\ref{ic_mixed.1}) we get
\begin{equation}
\langle N_+(t)\rangle= \sum_{i=1}^N \langle \theta(Z_{it}+x_{i0})\rangle
= \frac{1}{2}\, \sum_{i=1}^N {\rm erfc}\left(-\frac{x_{i0}}{\sqrt{2\, c(t,t)}}\right)\, ,
\label{ic_npav.1}
\end{equation}
where we used Eq. (\ref{theta_int.1}). Substituting Eq. (\ref{ic_npav.1})
in Eq. (\ref{ic_mixed.1}) gives
\begin{equation}
C_{\rm ic}(t_1,t_2)= \frac{1}{4}\,\sum_{ij} 
\overline{{\rm erfc}\left(\frac{-x_{i0}}{\sqrt{2\, c(t_2,t_2)}}\right)\,{\rm erfc}
\left(\frac{-x_{j0}}{\sqrt{2\, c(t_1,t_1)}}\right) }-\overline{{\rm erfc}
\left(\frac{-x_{i0}}{\sqrt{2\, c(t_2,t_2)}}\right)}\,\, 
\overline{{\rm erfc}\left(\frac{-x_{j0}}{\sqrt{2\, c(t_1,t_1)}}\right) }\, .
\label{ic1}
\end{equation} 
It is convenient to introduce the empirical density field of the initial condition
\begin{equation}
\rho_0(x) = \sum_{i=1}^N \delta(x-x_{i0})\, .
\label{ic_emp.1}
\end{equation}
Then using $\int f(x)\, \rho_(x)\, dx= \sum_{i=1}^N f(x_i)$ for arbitrary function $f(x)$,
we can rewrite Eq. (\ref{ic1}), in the large $N$ limit, as
\begin{equation}
C_{\rm ic}(t_1,t_2)= \frac{1}{4}\,\int_{-\infty}^0 dx\,\int_{-\infty}^0  dy\,
\left[\, \overline{\rho_0(x)\,\rho_0(y)}-\rho^2\,\right]\,
{\rm erfc}\left(-\frac{x}{\sqrt{2 c(t_2,t_2)}}\right)\,{\rm erfc}\left(
-\frac{y}{\sqrt{2 c(t_1,t_1)}}\right)\, ,
\label{ic_cic.1}
\end{equation}
where we assumed that the average initial density
$\overline{\rho_0(x)}=\overline{\rho_0(y)}=\rho$ is independent of space, i.e,
the density is uniform. 

To compute the double integral over semi-infinite space in Eq. (\ref{ic_cic.1}),
we then make use of a trick in Ref.~\cite{BJC22}. Let us assume that the
initial configuration with uniform density $\rho$ was spread over the full line
and then the particles are erased over the positive half. The density correlations
between particles on the left half then is unaffected by the erasure, provided
the correlations are short-ranged. Working then on the full line and
assuming translational invariance in the initial condition,
the connected part of the density-density correlation can be conveniently 
represented in Fourier space as
\begin{equation}
\overline{\rho_0(x)\,\rho_0(y)}-\rho^2 =\rho\, \int_{-\infty}^{\infty} \frac{dq}{2\pi}\, S(q)\,
e^{-i\, q\, (x-y)}\, ,
\label{ic_sf.1}
\end{equation}
where $S(q)$ is the structure factor. Substituting Eq. (\ref{ic_sf.1})
in Eq. (\ref{ic_cic.1}) and changing $x\to -x$ and $y\to -y$ for convenience,
one arrives at  
\begin{equation}
C_{\rm ic}(t_1,t_2)= \frac{\rho}{8\pi }\, \int_0^\infty dx\, \int_0^{\infty} dy\,
\int_{-\infty}^{\infty} dq\, S(q)\, e^{i\, q\, (x-y)}\, 
{\rm erfc}\left(\frac{x}{\sqrt{2 c(t_2,t_2)}}\right)\,{\rm erfc}
\left(\frac{y}{\sqrt{2 c(t_1,t_1)}}\right) \, .
\label{ic_cic.2}
\end{equation}

To proceed further, we make the change of variables
\begin{equation} 
x= u\, \sqrt{2}\,\left(c(t_1,t_1)\,c(t_2,t_2)\right)^{\frac{1}{4}}\,; \quad
y= v\, \sqrt{2}\,\left(c(t_1,t_1)c(t_2,t_2)\right)^{\frac{1}{4}}; \quad
{\rm and}\quad q= \frac{p}{\sqrt{2}\, 
\left(c(t_1,t_1)\,c(t_2,t_2)\right)^{\frac{1}{4}}}\, ,
\label{ic_cov.1}
\end{equation}
to obtain
\begin{equation}
C_{\rm ic}(t_1,t_2)= \frac{\rho}{8\,\pi}\,b(t_1,t_2)\,
\int_0^\infty du \int_0^{\infty}dv\, \int_{-\infty}^{\infty} dp\,
 S\left(\frac{p}{b(t_1,t_2)}\right)\,e^{i\,p\,(u-v)}\,
{\rm erfc}\left(\frac{u}{a(t_1,t_2)}\right)
{\rm erfc}\left(v\,a(t_1,t_2)\right)\, ,
\label{ic_cic.3} 
\end{equation}
where 
\begin{equation}
a(t_1,t_2)=\left(\frac{c(t_2,t_2)}{c(t_1,t_1)}\right)^{\frac{1}{4}}\, \quad
{\rm and}\quad b(t_1,t_2)= \sqrt{2}\, \left(c(t_1,t_1)c(t_2,t_2)
\right)^{\frac{1}{4}}\, .
\label{ic_defa}
\end{equation}

Now, at late times when $t_1$ and $t_2$ are both large, we expect
$b(t_1,t_2)$ to be large also, and hence we can make the approximation
\begin{equation}
S\left(\frac{p}{b(t_1,t_2)}\right)\simeq S(0) =\alpha_{\rm ic},
\label{ic_S0}
\end{equation}
where $\alpha_{\rm ic}$ is the Fano-factor or compressibility
associated with the initial conditions~\cite{BJC22}. Consequently,
one can do the integral over $p$ in Eq. (\ref{ic_cic.3}) to
give a delta function and the expression of $C_{\rm ic}(t_1,t_2)$ simplifies to
\begin{eqnarray}
C_{\rm ic}(t_1,t_2)&\approx & 
\frac{\alpha_{\rm ic}\,\rho\, b(t_1,t_2)}{4}\,\int_0^\infty du\, 
{\rm erfc}\left(\frac{u}{a(t_1,t_2)}\right)\,{\rm erfc}\left(u\, a(t_1,t_2)\right) 
\nonumber \\
&=& \frac{\alpha_{\rm ic}\,\rho\,b(t_1,t_2)}{4\,\sqrt{\pi} }\,
\left[\frac{1+ a(t_1,t_2)^2 -\sqrt{1+ a(t_1,t_2)^4}}{a(t_1,t_2)}\right]\,  \\
&=& {\frac{\alpha_{\rm ic}\,\rho}{2 \sqrt{2\,\pi}} \left( \sqrt{c(t_1, t_1)} + \sqrt{c(t_2,t_2)} - \sqrt{c(t_1,t_1)+c(t_2,t_2)}\right)} \;,
\label{ic_cic.4}
\end{eqnarray}
{where, in the last equality, we have used the expressions for $a(t_1,t_2)$ and $b(t_1, t_2)$ given in (\ref{ic_defa}).}
At equal times, we get
\begin{equation}
C_{\rm ic}(t,t)\approx \frac{\alpha_{\rm ic}\,\rho\, (\sqrt{2}-1)}{2\sqrt{\pi} }\,
\sqrt{c(t,t)}\, .
\label{ic_eqt.1}
\end{equation} 
The result in Eq. (\ref{ic_cic.4}) is
one of our main results in this section, which
demonstrates that
$C_{\rm ic}(t_1,t_2)$ does depend on the correlations in the initial condition,
but only through the Fano factor $\alpha_{\rm ic}$. Consequently, the 
annealed two-time correlation function in Eq. (\ref{ic_decom.1}), i.e.,
$C_{\rm an}(t_1,t_2)= C_{\rm qu}(t_1,t_2)+C_{\rm ic}(t_1,t_2)$ also
depends on the initial correlations through the Fano factor $\alpha_{\rm ic}$. {Note that for uncorrelated initial conditions we simply have 
$\alpha_{\rm ic} = 1$. This follows from the fact that, for uncorrelated (i.e. Poissonian) initial conditions, one has $\overline{\rho_0(x)\,\rho_0(y)}-\rho^2 = \rho \, \delta(x-y)$. Hence from Eq. (\ref{ic_sf.1}), we see that  $\alpha_{\rm ic} = S(0) = 1$. Therefore, by substituting the result in Eq.~(\ref{ic_cic.4}) (setting $\alpha_{\rm ic} = 1$) in Eq.~(\ref{ic_decom.1}), and using the result for the quenched correlation function in Eq. (\ref{ic_qu.4}), one recovers the result obtained in Eq.~(\ref{la11}) in the previous section.}

Let us now consider the specific example of fBM with correlator $c(t_1,t_2)$
given in Eq. (\ref{fbm_corr.1}). For this case we get from Eq. (\ref{ic_defa})
\begin{equation}
a(t_1,t_2)= \left(\frac{t_2}{t_1}\right)^{H/2}\, \quad {\rm and} \quad
b(t_1,t_2)= 2\, \sqrt{D_H}\, (t_1\, t_2)^{H/2}\, .
\label{ic_ab.1}
\end{equation}
Consequently, Eq. (\ref{ic_cic.4}) gives
\begin{equation}
C_{\rm ic}(t_1,t_2)\approx \frac{\alpha_{\rm ic}\, \rho\, \sqrt{D_H}}{2\, \sqrt{\pi}}\,
\left[t_1^H +t_2^H -\sqrt{t_1^{2H}+t_2^{2H}}\right]\, .
\label{ic_fbm.1}
\end{equation}
At equal times, one gets
\begin{equation}
C_{\rm ic}(t,t)\approx \frac{\alpha_{\rm ic}\,\rho\, (2\, 
D_H)^{\frac{1}{2}}\,(\sqrt{2}-1)}{2\,\sqrt{\pi} }\,t^H\, ,
\end{equation}
which agrees, for the Brownian case $H=1/2$, with
Ref.~\cite{BJC22}.

Finally, the fluctuations in the instantaneous current $j(0,t)$ 
coming from randomness in the initial conditions can 
also be be computed. Using the result in Eq. (\ref{ic_fbm.1}) one finds
\begin{equation}
\overline{\langle j(0,t_2)\rangle\langle j(0,t_1)\rangle}- 
\overline{\langle j(0,t_2)\rangle}\ 
\overline{\langle j(0,t_1)\rangle}\Big|_{\rm ic}=
\frac{\partial^2}{\partial t_2\partial t_1} C_{\rm ic}(t_1,t_2) \approx 
\frac{\alpha_{\rm ic}\, \rho\, H^2\, \sqrt{D_H}}{2\,\sqrt{\pi}}\frac{t_1^{2H-1} 
t_2^{2 H-1}}{\left(t_1^{2H} +t_2^{2H}\right)^\frac{3}{2}}\, ,
\label{ic_jj.1} 
\end{equation}
which, at equal times, is given by
\begin{equation}
\overline{\langle j(0,t)\rangle^2}- 
\left[\,\overline{\langle j(0,t)\rangle}\,\right]^2\Big|_{\rm ic}\approx
\frac{\alpha_{\rm ic}\, \rho\, H^2\, \sqrt{D_H}}{2\, \sqrt{2\pi} }\, t ^{H-2}.
\label{ic_jj.2}
\end{equation}
Thus, interestingly, the variance characterizing the current fluctuations, 
due to the correlations in the initial conditions, decays as a power 
law $\sim t^{-(2-H)}$ at late times for fBM.

\section{Conclusions}
\label{Summary}

We have studied the effusion of a general one dimensional stochastic process started 
on the negative half line into a target region to the right of the origin. For 
general stochastic processes, {when the initial distribution of the number of particles in a given
interval on the negative side is Poissonian, the number of particles in the target region remains Poissonian at all times -- albeit with a time-dependent parameter.} Interestingly, the joint distribution 
of the particle number in the region at two different times is given by a bivariate 
Poisson distribution. We have explicitly computed the parameters of this bivariate 
Poisson distribution for general zero mean Gaussian processes when the target region 
is the positive real axis. In particular we have given results for fractional 
Brownian motion and thermalized underdamped Brownian motion.

We have also analyzed both the annealed and the quenched two-time correlation functions 
of the number of particles in the positive real line.
In addition to the number of particles on the positive half line (or equivalently
the total integrated current through the origin up to time $t$), we
have also studied the fluctuations of the instantaneous current $j(0,t)$
through the origin at time $t$. While
this was well known for Brownian motion, we were able to 
generalize it to the case of general Gaussian processes where the current 
cannot necessarily be defined via a Fokker-Planck equation. 
Moreover we have also studied in detail how the correlations present in the
initial condition, captured by the Fano factor or compressibility
in the initial condition,  affects the two-time correlation functions.
Results previously obtained for the variance of the particle number at a single time for 
Brownian particles have been generalized to two-time quantities and for general Gaussian 
processes in our work. Finally, though we have focussed on the 
effusion problem, 
the results here can be extended to study generalizations of single file 
diffusion or other stochastic processes with reflection, via the relation in Eq. (\ref{link}) {which allows to obtain informations on the position $Y(t)$ at time $t$ of a tagged particle. In particular, using the fact that 
$N_+(t)$ and $N_-(t)$ are independent and identically distributed, this relation (\ref{link}) gives}
\begin{equation}
\overline{\langle Y(t)\rangle} =0 \;,
\end{equation}
and
\begin{eqnarray}
\overline{\langle Y(t_1) Y(t_2)\rangle}& =& \frac{1}{\overline{\rho}^2}\overline{\langle (N_+(t_1) - N_-(t_1))(N_+(t_2) - N_-(t_2))\rangle}=\frac{2}{\overline{\rho}^2}[\overline{\langle N_+(t_1) N_+(t_2)\rangle} - \overline{\langle N_+(t_2)\rangle} \ \ 
 \overline{ \langle N_+(t_1)\rangle}]\nonumber\\
 &=& \frac{2}{\overline{\rho}^2}C_{\rm an}(t_1,t_2) \;.
 \end{eqnarray}

For the annealed initial configuration, we have derived here
the full two-time joint distribution of $N_A(t)$.
Using our method, one can readily see how general $n$ point statistics can be treated in 
principle. However explicit results require the computation of the $n$-point 
correlation function of the indicator function $I_A(X(t))$ and closed analytical 
forms for such observables are {notoriously} difficult to obtain \cite{BMS13}. It would also be interesting to extend 
the results obtained here to the case where $A$ is a finite subset of the real line in order 
to understand number fluctuations in finite sized pore or channels in the spirit of 
the study in \cite{MAR21}. In \cite{VCCK08} the authors used the exact stochastic 
density equation \cite{DEA96} for noninteracting Brownian motions to derive the 
equilibrium Poisson statistics for this system. It would be interesting to see how 
their method can be adapted to this non-equilibrium effusion problem and explore the 
links with approaches using macroscopic fluctuation theory \cite{BDG01}.

\end{document}